\documentclass{article}
\usepackage[accepted]{icml2007}

\pdfoutput=0

\usepackage{epsf}
\usepackage{mlapa}
\usepackage{psfrag}
\usepackage{graphicx}
\usepackage{amsfonts,amsmath}

\newcommand{\mf}[1]{\mathbf{#1}}
\newcommand{\bq}{\begin{equation}}
\newcommand{\eq}{\end{equation}}
\newcommand{\bqn}{\begin{eqnarray}}
\newcommand{\eqn}{\end{eqnarray}}
\newcommand{\fr}[2]{\frac{#1}{#2}}
\newcommand{\bd}[1]{\mathbf{#1}}

\icmltitlerunning{Robust Mixtures in the Presence of Measurement
Errors}

\begin{document}

\twocolumn[ \icmltitle{Robust Mixtures in the Presence of
Measurement Errors}

 \icmlauthor{Jianyong Sun}{jxs@cs.bham.ac.uk}
 \icmladdress{School of Computer Science and School of Physics \& Astronomy,
              University of Birmingham, Birmingham, UK}
 \icmlauthor{Ata Kab\'{a}n}{axk@cs.bham.ac.uk}
 \icmladdress{School of Computer Science, University of Birmingham, Birmingham, UK}
 \icmlauthor{Somak Raychaudhury}{somak@star.sr.bham.ac.uk}
 \icmladdress{School of Physics \& Astronomy, University of Birmingham, Birmingham, UK}

\vskip 0.3in ]

\begin{abstract}
We develop a mixture-based approach to robust density modeling and
outlier detection for experimental multivariate data that includes
measurement error information. Our model is designed to infer
atypical measurements that are not due to errors, aiming to retrieve
potentially interesting peculiar objects. Since exact inference is
not possible in this model, we develop a tree-structured variational
EM solution. This compares favorably against a fully factorial
approximation scheme, approaching the accuracy of a Markov-Chain-EM,
while maintaining computational simplicity. We demonstrate the
benefits of including measurement errors in the model, in terms of
improved outlier detection rates in varying measurement uncertainty
conditions. We then use this approach in detecting peculiar quasars
from an astrophysical survey, given photometric measurements with
errors.
\end{abstract}

\section{Introduction}

The goal in robust unsupervised data modeling is to capture the
structure of the typical observations while dealing with atypical or
outlying observations in an automated manner. Outliers can occur for
various reasons, such as measurement errors or the existence of
peculiar objects in a data set. If atypical observations exist and
are not properly dealt with, they lead to biases in the parameter
estimates and poor generalization of the structure inferred from the
data. Therefore, a great deal of effort has been invested into
modifying existing unsupervised methods to provide them with
robustness properties. In the statistics and statistical machine
learning communities, the Student t-distribution was put forth and
adopted as a robust building block, for clustering
\cite{Outlier:peel00,Outlier:sve&bis}, visualization
\cite{Outlier:vel&lis} and robust projections \cite{Outlier:arc&del}. The
t-distribution has heavy tails, hence it gives non-zero probability
to observations that are far away from the bulk of the density.

Apart from the issue of robustness of the parameter estimates, the
ability of detecting outliers is of special interest in certain
scientific areas such as in Astrophysics \cite{DJO}, where finding
peculiar objects from large archives of multi-wavelength astronomical
images provide a
unique means of identifying candidates of possibly new types of
objects that deserve more detailed follow-up study (e.g. using
spectroscopy). However, a bottleneck already anticipated in
\singleemcite{DJO} is `the likely overabundance of interesting objects
found' -- the interpretation and understanding of which will
necessitate costly detailed analysis. Indeed, not every atypical 
observation is truly interesting. One reason for this lies in
measurement errors resulting from uncertainties in
instrument calibration and physical limitations of devices and
experimental conditions. These
errors are typically carefully recorded in the case of scientific
data and are available. Yet, most existing data analysis
methods have no natural ways of taking these into consideration. 
In turn, neglecting the error information
holds the risk of compromising the accuracy with which
\emph{genuine} outliers can be detected, since there is nothing to
prevent us from confusing erroneous measurements with potentially
interesting rare or peculiar ones.

In classical statistics, models known as `errors in variables'
exist, such as the total least square approach for robust regression
\cite{huffel02}. Probabilistic approaches able to propagate
uncertainty have also started to appear recently \cite{LAWRENCE,ROD}
for certain problems, and their benefits have been convincingly
demonstrated. However we are aware of no work on including knowledge
of observational errors specifically for unsupervised robust density
modeling. Due to the importance of this issue in scientific data
mining, this paper makes an attempt to fill in this gap.

\section{Robust mixtures for data with errors}
\label{model}

%
%
Consider a data set in which each individual measurement is an estimate 
of the form $t_{in} \pm \sqrt{s_{in}}$,
where $n=1,...,N, i=1,...,d$, $N$ is the number of object instances
and $d$ is the number of features. It is conceptually justified to 
assume that the error associated with these individual measurements 
is normally distributed \cite{Taylor}. 
Organizing the square of errors into diagonal
matrices $\bd{S}_n$, for each measured $d$-dimensional 
data point $\bd{t}_n$, the following heteroscedastic noise model can
be written. {\small\begin{equation} p(\mathbf{t}_n|\mathbf{w}_n) =
{\cal N}(\mathbf{t}_n|\mathbf{w}_n, \mf{S}_n);\label{noise}
\end{equation}}where ${\cal N}(\mathbf{t}_n|\mathbf{w}_n, \mf{S}_n)$ denotes the
normal distribution with unknown mean $\mathbf{w}_n$ and known
diagonal covariance matrix $\mf{S}_n$.

The unknown mean values $\bd{w}_n$ represent the clean, error-free
version of the data. Since these cannot be observed directly, we
will treat them as latent variables. The genuine outliers, which we
are interested in, must be those of the density of $\bd{w}$ rather
than those of the density of $\bd{t}$. We will therefore model the
hidden clean density as a robust mixture of Student $t$-distributions
(MoT)\footnote{The instance indices $n$ will be
dropped for convenience, whenever their presence is obvious from the
context.}:
{\small\begin{equation} p(\mf{w}) = \sum\limits_{k=1}^K \pi_k
p(\mf{w}|k)\label{mixture}\end{equation}}where $p(\mf{w}|k)$ is the Student
$t$-density. By the use of $t$-densities, we make no assumptions on the distribution
of outliers. Outliers are instances outside the high density `cluster'
regions.
{\small\begin{equation}
S_t(\mf{w}|\mu_k,\mf{\Sigma}_k,\nu_k)
=\frac{\mf{\Gamma}(\fr{\nu_k+d}{2})|\mf{\Sigma}_k|^{-1/2}
\mf{\Gamma}(\fr{\nu_k}{2})^{-1}(\nu_k\pi)^{-\fr{d}{2}}} {\left(1
+\frac{(\mf{w}-\mu_k)^T\mf{\Sigma}_k^{-1}(\mf{w}-\mu_k)}{\nu_k}
\right)^{\fr{\nu_k+d}{2}}} \nonumber\end{equation}}

As noted in \singleemcite{Outlier:liu&rub}, with the introduction of an
auxiliary hidden variable $u$, the $t$-distribution can be re-written
as a convolution of a Gaussian with a Gamma placed on its 
precisions,{\small \bq  S_t(\mathbf{w}|\mu,\mf{\Sigma}, \nu) =
\int\limits_{0}^{\infty} {\cal N}(\mathbf{w}|\mu,
\fr{\mf{\Sigma}}{u}){\cal
G}(u|\frac{\nu}{2},\frac{\nu}{2})du;\label{clean} \eq}where $\cal G$
is the Gamma density, ${\cal G}(u|a,b) = b^a
u^{a-1}\frac{\exp({-bu})}{\mf{\Gamma}(a)}.$ This re-writing has been
exploited for developing an exact ML
estimation algorithm for the MoT model \cite{Outlier:peel00}. 


In our model, the distribution of the observed data $\mathbf{t}$ can
be obtained by integration over $\mathbf{w}$. According to Eqs.
(\ref{noise}), (\ref{mixture}) and (\ref{clean}), we have:
{\scriptsize\bq p(\mathbf{t}) = \sum_{k} \pi_k \iint{\cal
N}\left(\mathbf{t}|\mathbf{w}, \mf{S}\right) {\cal
N}\left(\mathbf{w}|\mu_k,\fr{\mf{\Sigma}_k}{u}\right) {\cal
G}\left(u|\fr{\nu_k}{2},\fr{\nu_k}{2}\right)dud\mathbf{w}\label{mot_error}
\eq}Thus, given a set of training data ${\cal Y}= (\mathbf{t}_1,
\cdots, \mathbf{t}_N)$, the complete probability model of the
observed variable $\mathbf{t}$ and the latent variables
$\mathbf{w},u,z$ will have the following factorized form:{\small
\bqn {\cal L}_C &=& \prod_{n}\prod_k [p(\mathbf{t}_n|\mathbf{w}_n)
p(\mathbf{w}_n|u_n,z_n=k)]^{\delta(z_n=k)}\times\nonumber\\
&&\prod_{n}\prod_k[p(u_n|z_n=k)p(z_n=k|\pi)]^{\delta(z_n=k)}\label{complete}
\eqn}where $\delta(\cdot)$ is the Kronecker delta. The plate diagram
representation of this model is shown on the right-hand plot of
Fig. \ref{graph}, along with that of the MoT model.
\begin{figure}[htbp]
\begin{center}
[This figure is available upon request from author]
\caption{Plate diagrams of MoT (left), and the proposed model
(right).}\label{graph}
\end{center}
\vspace{-30pt}
\end{figure}

\section{A structured variational EM solution}\label{gem}

Since the integration in Eq. (\ref{mot_error}) is not tractable, we
develop a generalized EM (GEM) algorithm (see e.g. \singleemcite{hogg05}),
with approximate E-step. In general terms, for each data point
$\mf{t}_n$, its log-likelihood can be written as follows, for any distribution $q$:
{\small
\bqn\log p(\mathbf{t}_n|\theta) &=&  \int q(h_n)\log \frac{p(h_n,
\mathbf{t}_n|\theta)}{q(h_n)}
\frac{q(h_n)}{p(h_n|\mathbf{t}_n,\theta)}dh_n \nonumber\\
&\geq& \int q(h_n) \log \frac{p(h_n, \mathbf{t}_n|\theta)}{q(h_n)}
dh \equiv {\cal F}(\mf{t}_n|q,\theta)\label{gemeq} \nonumber\eqn }where $q$
is the free-form variational family (or variational posterior),
$\cal F$ is called the variational free energy function, $h_n$ is the
set of latent variables associated with $\mf{t}_n$, and $\theta$ is
the set of parameters of the model. In our case, $h_n = (z_n,
\mathbf{w}_n, u_n)$ and $\theta = (\{\mu_k\}, \{\mf{\Sigma}_k\},
\{\nu_k\}, \pi)$. The log-likelihood of the given data set $\cal Y$
is then lower bounded by the free energy: 
{\small \bq \log p({\cal Y}|\theta) =
\sum\limits_n \log p(\mf{t}_n|\theta) \geq \sum\limits_n {\cal
F}(\mf{t}_n|q(h_n), \theta)\label{bound}\eq} 
In the E-step of the $(k+1)$-th iteration of a GEM algorithm, we maximize
$\cal F$ w.r.t the variational distribution $q$ while fixing the
parameters in the $k$-th iteration, $\theta^{k}$:
{\small\begin{equation} q^{k+1}(h_n) = \arg \max_{q} {\cal
F}(\mathbf{t}_n|q, \theta^{k}).
\end{equation}}
In the M-step, we maximize Eq. (\ref{bound}) w.r.t the parameters
$\theta$ to obtain the new parameter values $\theta^{k+1}$:
{\small\begin{equation} \theta^{k+1} = \arg \max_{\theta}
\sum\limits_n {\cal F}(\mathbf{t}_n|q^{k+1}, \theta).
\end{equation}}

\subsection{Tree-structured variational
distribution}\label{algorithm}

Some tractable form needs to be chosen for $q$. 
The most common choice is a fully factorial form
\cite{jordan99}. In our case, this would be
$q(\bd{w},u,z)\equiv q(\bd{w})q(u)q(z)$. In the context of robust
mixtures, fully factorial variational posterior distributions have
been employed in \cite{Outlier:sve&bis}, though with a slightly different
model specification.
Let us observe, however,
that under our model definitions, it is feasible to keep some of the
posterior dependencies by choosing the following tree-structured
variational distribution: {\small$$
q(\mathbf{w},u,z=k)=q(z=k)q(\mathbf{w}|z=k)q(u|z=k)$$}Structured
variational distributions have been used previously in the context
of various other latent variable models
\cite{Str:gei&mee,Str:bis&win} and have been found more accurate
compared to the fully factorial choice. Yet, their use is still not
as popular as it could be. In the following, we denote
$q(z=k)$ by $q(k)$, $q(\mf{w}|z=k)$ by
$q(\mf{w}|k)$ and $q(u|z=k)$ by $q(u|k)$. Also, expectations w.r.t.
$q(\bd{w}|z=k)$ will be denoted by $\langle . \rangle_{\bd{w}|k}$
and similarly, those w.r.t. $q(u|k)$ by $\langle . \rangle_{u|k}$,
and those w.r.t. the joint $q(\bd{w},u|k)= q(\bd{w}|k)q(u|k)$ by
$\langle . \rangle_{\bd{w},u|k}$.

\subsection{Deriving the GEM algorithm}

The free energy function ${\cal F}(\mathbf{t}|q,\theta)$ can be
evaluated as {\small\begin{eqnarray} && {\cal
F}(\mathbf{t}|q,\theta)= \sum_k q(k)\left[\langle \log
p(\mathbf{t,w},u,k)\rangle_{\mathbf{w},u|k}\right]+
H(q)\nonumber\eqn}where $H(q)$ is the entropy of the variational
distribution: {\small $ H(q) = - \sum_k q(k)\left[\langle \log
\left(q(u|k)q(\mf{w}|k)q(k)\right)\rangle_{\mf{w},u|k}\right]$.}
Defining {\small $A_{\mathbf{t},k} = \langle \log
p(\mathbf{t,w},u,k)\rangle_{\mathbf{w},u|k} - \langle \log
q(u|k)\rangle_{u|k}- \langle \log
q(\mathbf{w}|k)\rangle_{\mathbf{w}|k}$} then we have: {\small\bq
{\cal F}(\mathbf{t}|q,\theta)= \sum_k
q(k)\left[A_{\mathbf{t},k}-\log q(k)\right].\label{lowerbound}\eq}

\subsubsection{Variational E-step}

Now, in order to find the optimal functional form of the posterior
distribution terms, we take functional derivatives of ${\cal
F}(\mf{t}|q,\theta)$ w.r.t. the terms of $q$, i.e.
$q(\mathbf{w}|k)$, $q(u|k)$ and $q(k)$ respectively, and equate
these to the identically null function. We obtain the following:
{\small\begin{eqnarray} q(\mathbf{w}|k) &=& \frac{\exp\langle \log
\left[p(\mf{t}|\mf{w})p(\mf{w}|u,k)\right] \rangle_{u|k}}{\int
\exp\langle \log \left[p(\mf{t}|\mf{w})p(\mf{w}|u,k)\right]
\rangle_{u|k} d\mf{w}}\label{qwk}\\
q(u|k) &=& \frac{\exp\langle\log\left[p(\mf{w}|u,k)p(u|k)\right]
\rangle_{\mf{w}|k}}{\int
\exp\langle\log\left[p(\mf{w}|u,k)p(u|k)\right]
\rangle_{\mf{w}|k} du}\label{quk}\\
q(k) &=&
\frac{\exp(A_{\mathbf{t},k})}{\sum_{k'}\exp(A_{\mathbf{t},k'})}\label{qk}
\end{eqnarray}}It can be seen that $q(\mf{w}|k)$ and $q(u|k)$ depend only on 
variables in their Markov blanket. However, the
distribution $q(k)$ depends on all other variables in the graph.
Conveniently, the quantities required for computing Eq. (\ref{qk}) will
be available from the computations that are needed for evaluating the
free energy function --- which in turn is useful for monitoring the
convergence of the GEM iterations.

Due to the conjugacy properties of the distributions we used, and
after simplification, we now can obtain $q$ analytically. Let us
define: {\small\bq \mf{\Sigma}_{\mathbf{w}|k} = \mf{S}\left[
\frac{\mf{\Sigma}_k}{\langle u\rangle_{u|k} } + \mf{S}\right]^{-1}
\frac{\mf{\Sigma}_k}{\langle u\rangle_{u|k}} \label{trucov}\eq}
{\small \bq  \langle \mathbf{w}\rangle_{k} =
\mf{\Sigma}_{\mathbf{w}|k}\left( \langle u\rangle
_{u|k}\mf{\Sigma}_k^{-1}\mu_k + \mf{S}^{-1}\mathbf{t} \right)
\label{trumu}\eq} { \small\bq  a_k = \frac{\nu_k+d}{2};\;\;\;\;\;\;\;\;\;\; b_k =
\fr{\nu_k+C_k}{2} \label{vb} \eq} where {\small\bq  C_k =
\left(\langle
\mathbf{w}\rangle_{k}-\mu_k\right)^T\mf{\Sigma}_k^{-1}\left(\langle
\mathbf{w}\rangle_{k}-\mu_k\right)+\mf{Tr}\left(\mf{\Sigma}_k^{-1}
\mf{\Sigma}_{\mf{w}|k}\right).\label{vc} \eq} 
Then we have:{\small
\bq q(\mathbf{w}|k) ={\cal N}(\mathbf{w}|\langle
\mathbf{w}\rangle_{k}, \mf{\Sigma}_{\mathbf{w}|k});\ \ q(u|k) ={\cal
G}(u|a_k, b_k).\label{quk1}\eq}


\subsubsection{The variational likelihood bound}

$A_{\mathbf{t},k}$ can be evaluated as follows: {\small
\begin{eqnarray} A_{\mathbf{t},k} &=& \langle\log
p(\mathbf{t}|\mathbf{w})\rangle_{\mathbf{w}|k} + \langle\log
p(u|k)\rangle_{u|k}+ \log \pi_k+ \nonumber\\
&&\langle\log p(\mathbf{w}|u,k)\rangle _{\mathbf{w},u|k}-
\langle\log q(\mathbf{w}|k)\rangle _{\mathbf{w}|k} -\nonumber\\
&& \langle\log q(u|k)\rangle _{u|k}\nonumber\\ &=&
Q_1+Q_2+Q_3+Q_4+Q_5+Q_6\label{atnk}
\end{eqnarray}}
where 
{\small 
\begin{eqnarray*} Q_1 &=& -\frac{d}{2}\log(2\pi) - \frac{1}{2}\log
|\mf{S}|-\fr{1}{2}\mf{Tr}(\mf{\Sigma}_{\mf{w}|k}\mf{S}^{-1})
\end{eqnarray*}
\begin{eqnarray*}
&& - \frac{1}{2}\left[(\langle\mathbf{w}\rangle_{k} -
\mf{t})^T\mf{S}^{-1}(\langle\mathbf{w}\rangle_{k} -
\mf{t})\right];\\ 
Q_2 &=& (\frac{\nu_k}{2}-1)\langle\log
u\rangle_{u|k} - \frac{\nu_k}{2}\fr{a_k}{b_k} \\
&& + \frac{\nu_k}{2} \log(\frac{\nu_k}{2})-\log
\mf{\Gamma}(\frac{\nu_k}{2});\\ 
Q_3 &=&-\frac{d}{2}\log(2\pi) - \frac{1}{2}\log|\mf{\Sigma}_k| +
\frac{d}{2}\fr{a_k}{b_k} \\
&&- \frac{1}{2}\fr{a_k}{b_k} \left[(\langle \mathbf{w}\rangle_{k}-
\mu_k)^T\mf{\Sigma}_k^{-1}(\langle
\mathbf{w}\rangle_{k}- \mu_k)\right]\\
&& - \frac{1}{2}\fr{a_k}{b_k}
\mf{Tr}(\mf{\Sigma}_{\mf{w}|k}\mf{\Sigma}_k^{-1});\\
Q_4 &=& \log \pi_k;\\
Q_5 &=& \frac{d}{2} + \frac{d}{2}\log(2\pi)+\frac{1}{2}\log |\mf{\Sigma}_{\mathbf{w}|k}|;\\
Q_6 &=& -[(a_k-1)\langle\log u\rangle_{u|k} +a_k\log b_k-a_k - \log
\mf{\Gamma}(a_k)];
\end{eqnarray*}
}and where $\langle \log u\rangle_{u|k} = \psi(a_k) - \log b_k$ and
$\psi(\cdot)$ is the di-gamma function.

In summary, given a data $\cal Y$, the log likelihood bound is computed 
cf. Eq. (\ref{lowerbound}) as the following:
{\small \begin{equation}
{\cal F} = \sum_{n}\sum_{k}q(z_n=k)\left[A_{\mathbf{t}_n, k}-\log
q(z_n=k)\right]\label{Q}
\end{equation}}where $A_{\mathbf{t}_n,k}$ is computed as in Eq. (\ref{atnk}) for
each data point $\mf{t}_n$. Eq. (\ref{Q}) is useful to monitoring
the convergence.

\subsubsection{M-step}

The parameter re-estimates are obtained by solving the
stationary equations of $\cal F$ w.r.t $\mu_k$, $\mf{\Sigma}_k$ and
$\pi_k$, which yields: 
{\small
\begin{eqnarray} \mu_k &=& \frac{\sum_{n=1}^{N}q(z_n=k)\langle
u_n\rangle_{u_n|k} \langle \mf{w}_n \rangle_{\mathbf{w}_n|k}}
{\sum_{n=1}^{N}q(z_n=k)\langle u_n\rangle_{u_n|k} } \label{mu}\\
\mf{\Sigma}_k &=& \frac{\sum_{n=1}^{N}q(z_n=k)\langle
u_n\rangle_{u_n|k}
\tilde{\mf{\Sigma}}_{n,k}}{\sum_{n=1}^{N}q(z_n=k)} \label{cov}\\
\pi_k &=& \frac{1}{N}\sum_{n=1}^{N}q(z_n=k)\label{pk}
\end{eqnarray}}where {\small $\tilde{\mf{\Sigma}}_{n,k}
=\left[(\mu_k-\langle\mf{w}_n\rangle_{k})
(\mu_{k}-\langle\mf{w}_n\rangle_{k})^{T} +
\mf{\Sigma}_{\mathbf{w}_n|k}\right]$}. Finally, $\nu_k$ is re-estimated
by solving the following non-linear equation. {\small
\begin{eqnarray*}
\sum_n q(z_n=k)[\log(\frac{\nu_k}{2})+1+\langle\log u_n
\rangle_{k}-\fr{a_{nk}}{b_{nk}}-\psi(\frac{\nu_k}{2})]=0.\label{nuk}
\end{eqnarray*}}

\subsection{Scaling}
Considering the time complexity of the algorithm, per iteration,
computing the posterior mean and covariance ($\langle
\mf{w}\rangle_{k}$ and $\mf{\Sigma}_{\mf{w}|k}$) for each data point
$\mf{t}$ takes ${\cal O}(d^3K)$ operations. The computation of the
parameters of $q(u|k)$, $a_k$ and $b_k$ take ${\cal O}(d^3K)$, and
the responsibility $q(k)$ needs ${\cal O}(d^3K)$ time as well.
In total, this is ${\cal O}(d^3KN)$. 
For comparison, the maximum likelihood estimation of MoT \cite{Outlier:peel00}
takes ${\cal O}(d^3K)$ to compute $p(u|k,\mf{t})$ and ${\cal O}(d^3K)$ to compute
$p(k|\mf{t})$, which totals a complexity of ${\cal O}(d^3KN)$ --- same as that of
proposed algorithm. Moreover, using a full factorial approximation in our model
also results in the same theoretical complexity per iteration. 
So the only extra burden of our proposed method is the computation of
the posterior mean and covariance of the clean data
$\mf{w}$. The most expensive operation appears to be the matrix
inversion, however, it should be noted, this is only required when
$\bd{\Sigma}_k$ are modeled as a full covariances, which is feasible
in relatively low-dimensional problems ($d \ll N$). If this model was
to be used on high dimensional data, then a diagonal form
$\bd{\Sigma}_k$ would need to be taken
--- in which case the cubic operation is no longer required since
the matrices to be inverted become diagonal.

\subsection{Accommodating new data points}

Since the model is fully generative, it can also be applied to new,
previously unseen data from the same source. For a given test data
set, we need to calculate the posterior distributions of $\mf{w}_n$
and $u_n$ associated with each test point $\mf{t}_n$. To calculate
these, we fix the parameters $\mu_k$, $\mf{\Sigma}_k$ and $\pi_k$,
$1\leq k \leq K$ obtained from the training set and perform the
E-step iterations until convergence. This typically converges at
least an order of magnitude faster than the full training procedure.

\subsection{Determining the number of components}

To determine the number of mixture components, the minimum message 
length (MML) criterion \cite{figueiredo02} could be employed.
We derive a lower bound to MML, so
the optimal number of components is found by maximizing the
following criterion:
{\small \begin{eqnarray} 
{\cal L}(\theta, {\cal Y}) &=& -\fr{\hat{n}}{2}\sum\limits_{k:\pi_k
> 0}\log\left(\fr{N\pi_k}{12}\right) -
\fr{k_{nz}}{2}\log\left(\fr{N}{12}\right)\nonumber\\
&&-\fr{k_{nz}(\hat{n}+1)}{2}+\log p({\cal Y}|\theta)\label{mml}
\end{eqnarray}}where $p({\cal Y}|\theta)$ is the data log-likelihood,
$\hat{n}$ is the dimensionality of the parameters, $k_{nz}$ is the
number of non-zero-probability components. The free parameters involved
in the proposed algorithm are the means and the full covariance
matrices of ${\mathcal N}(\mf{w}_n|u_n,z_n=k)$.
Thus the dimensionality of the $k$-th parameter $\theta_k = (\mu_k,
\mf{\Sigma}_k)$, is $d + d(d-1)/2$.
We approximate the data likelihood as earlier, by Eq. (\ref{bound}).
Replacing this in (\ref{mml}), leads to maximizing:{\small \bqn {\cal
{L}}(\theta, {\cal Y}) &\geq& -\fr{\hat{n}}{2}\sum\limits_{k:\pi_k >
0}\log\left(\fr{N\pi_k}{12}\right) -
\fr{k_{nz}}{2}\log\left(\fr{N}{12}\right)\nonumber\\
&-&\fr{k_{nz}(\hat{n}+1)}{2} + \sum\limits_n {\cal F}(\mf{t}_n|q,\theta)
\equiv {\cal \tilde{L}}({\cal Y}|q,\theta).\nonumber\eqn}This maximization is
similar to the GEM presented in Section \ref{algorithm}
and algorithmically the only difference is in computing the mixing
proportions $\pi_k$ in the M-steps, which is now: {\small \bqn \pi_k
= \fr{\max\left\{0, \sum_{n=1}^N
q(z_n=k)-\fr{\hat{n}}{2}\right\}}{\sum_{j=1}^K\max\left\{0,
\sum_{n=1}^N q(z_n=j)-\fr{\hat{n}}{2}\right\}}\eqn} Of course, only
the non-zero-probability components of the mixtures will contribute
to $q(\mf{w}_n|z_n)$, $q(u_n|z_n)$ and $q(z_n)$.

\section{The Outlier Detection Criteria}\label{out}

Since we modelled the clean, error-free data by a mixture of 
$t$-distributions, we would expect that the model can find outliers
w.r.t the clean data, rather than the contaminated data. Following
\singleemcite{Outlier:peel00}, the posterior expectation of $u$ is interpretable as an
outlierness indicator. Using our posterior approximations described earlier
(i.e. $q(u,k)=q(u|k)q(k)$ to
approximate $p(u,k|\mathbf{t})$), then the variational expectation of $u$ will
be employed to infer outlierness. This is:
{\small \begin{equation} e
\equiv\sum_k q(k)\frac{\nu_k+d}{\nu_k+\mf{Tr}(\mf{\Sigma}_k^{-1}
\mf{\Sigma}_{\mathbf{w}|k})+\mf{\Delta}_{\mathbf{w}|k}^2}\label{outlier}
\end{equation}}where $a_k$ and $b_k$ are defined in Eq. (\ref{vb});
and {\small $\mf{\Delta}_{\mathbf{w}|k}^2 =
(\langle\mf{w}\rangle_{\mf{w}|k}-\mu_k)^T\mf{\Sigma}_k^{-1}
(\langle\mf{w}\rangle_{\mf{w}|k}-\mu_k)$}. Therefore, a data point
is considered to be an outlier if its corresponding $e$ value is
sufficiently small. 

In contrast, recall that for MoT, the outlier criterion value
\cite{Outlier:peel00} is {\small \begin{equation} e_{MoT} \equiv \sum_k
p(k|\mathbf{t}) \frac{\nu_k+d} {\nu_k +
(\mathbf{t}-\mu_k)^{T}\mf{\Sigma}_k^{-1}(\mathbf{t}-\mu_k)}\label{mot}
\end{equation}}
So we see that rather than instead of the Mahalanobis distance between 
the mean $\mu_k$ and the data $\mathbf{t}$ as in Eq. (\ref{mot}), 
the distance between the center $\mu_k$ and the expected value of the clean 
data $\mathbf{w}$ is present in (\ref{outlier}).

Further, it can easily be seen, for consistency, that in the limit of zero
observation error, our outlierness criterion reduces to that
of MoT. Indeed, whenever $\mf{S}=\textbf{0}$,
Eqs. (\ref{trucov}),(\ref{trumu}) and (\ref{vc}) can be written as:
{\small \begin{eqnarray} \mf{\Sigma}_{\mathbf{w}|k} = \textbf{0};&&
\langle\mf{w} \rangle_{\mathbf{w}|k} = \mathbf{t};\nonumber\\
C_k &=&
(\mathbf{t}-\mu_k)^{T}\mf{\Sigma}_{k}^{-1}(\mathbf{t}-\mu_k);\nonumber
\end{eqnarray}}Then replacing the above equations into Eqs. (\ref{quk}),
(\ref{mu}), (\ref{cov}) and (\ref{pk}), we can recover the
posteriors as: \bq q(u|k) = p(u|\mathbf{t},k); q(k) =
p(k|\mathbf{t});\nonumber \eq and so, the update formulas of the MoT
are recovered~\cite{Outlier:peel00}.

If the size of the measurement error $\mf{S}$ (we can measure
the size of $\mf{S}$ by its trace) is small, we expect the difference 
between $e$ and $e_{MoT}$ is relatively small too. However, as 
the size of the measurement error gets larger, the difference between
the two outlierness criteria becomes larger and consequently the ranking they produce
will be different. In particular, we can gain more insights and see the effects of 
a misspecification of the error by rewriting the data likelihood (\ref{mot_error}) by
integrating over $\mathbf{w}$: 
{\scriptsize \begin{equation}
p(\mathbf{t}_n)=\sum_{k} \pi_k \int{\cal
N}\left(\mathbf{t}_n|\mathbf{\mu}_k,\fr{\mf{\Sigma}_k}{u_n}+ \mf{S}_n\right) {\cal
G}\left(u|\fr{\nu_k}{2},\fr{\nu_k}{2}\right)du_n
\end{equation}}
The posterior expectations $\langle u_n\rangle$ are data instance-specific, 
ensuring the robustness of the parameter estimates, even if the errors (diagonals of $\mathbf{S}_n$)
are misspecified. However, this also implies that a data instance with an underestimated 
$\mathbf{S}_n$  gets picked as a false `interesting' outlier ($\langle u_n \rangle$ 
gets smaller). Clearly, if all errors are specified at zero, our model reduces to MoT and 
produces unwanted false detections.
\section{Experiments and results}\label{experiment}

To test the performance
of the proposed algorithm, first we experimentally assess
the accuracy of the structured factorization employed. Second, we
perform a set of controlled experiments on synthetic and
semi-synthetic data sets, in order to demonstrate the ability 
of detecting genuine outliers. Finally, we shall present a
real application of our approach in astronomy, for finding peculiar
(high redshift) objects from the SDSS quasar
catalogue \cite{vostat}.

\subsection{Synthetic data \& illustrative experiment}

A synthetic data set is constructed comprising of
error-free values sampled from a
mixture of three well separated Gaussians and a uniform distribution
simulates the presence of genuine outliers. Then we add Gaussian noise
to all points, to simulate
measurement errors, and apply our algorithm to the resulting
dataset. The aim is to
recover the genuine outliers (along with the density of
non-outliers), despite the Gaussian noise added. The leftmost plot
of Fig. \ref{type2} shows the error-free data, with the Gaussian
covariances of the clusters of non-outliers superimposed. Different markers are
used for points in different clusters and the
outliers are marked with stars. The central plot shows the
effect of simulating measurement errors. The marker sizes are
proportional to the size of errors. Notice that due to the
errors, some outliers appear closer to the main density regions while
some of the non-outliers `jump' away from the bulk of density. Thus
the measurement errors make the problem of recovering genuine
outliers much more challenging. The rightmost plots of Fig.
\ref{type2} shows the result of our estimation procedure described
earlier, superimposed over the data with errors.
\begin{figure}[htbp]
    \includegraphics[width = 3.4in, height=2.3in]{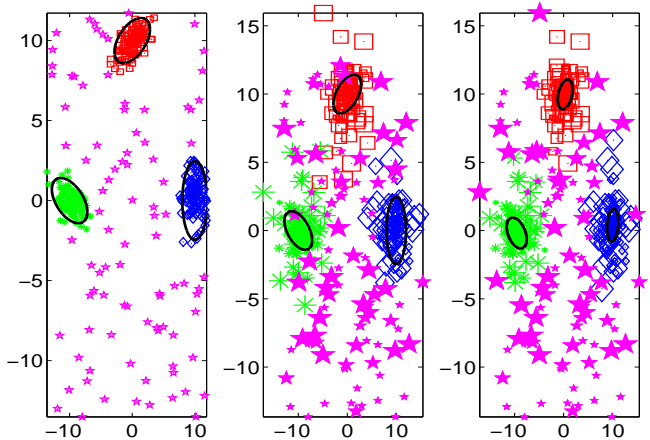}
    \caption{A synthetic data set with cluster structure and outliers.
    Left: Hidden error-free data with genuine outliers;
    Center: Data contaminated with measurement errors; Right: The estimated grouping and detected outliers.}
    \label{type2}
\end{figure}

\subsection{Comparison of alternative approximate EM methods}

Now, we test the accuracy of the structured variational EM method
developed, against a fully factorial variational EM for the same
model, and a Markov Chain EM (MCEM) realized through Gibbs sampling,
the latter being considered to represent the `ground truth'. Fig.
\ref{gibbs} shows the approximation of the log likelihood against
iterations in a run on the synthetic data set shown earlier. For
Gibbs sampling MCEM, $M=10,000$ samples were used for computing the
posterior estimates. The first 2000 samples were discarded as
burn-in. All algorithms were started from the same initial parameter
values.
\begin{figure}[htbp]
  \begin{center}
      \includegraphics[width = 3.0in]{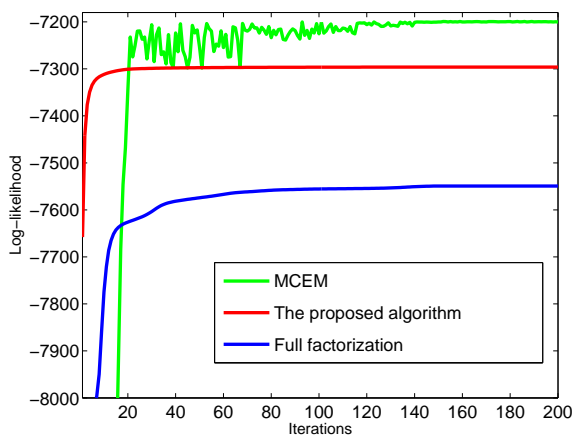}
\caption{The optimization process of alternative
approximate EM algorithms on the synthetic data set.}\label{gibbs}
  \end{center}
\end{figure}
As expected, MCEM is superior to variational methods, but at the
price of a heavy computational demand and difficulties in
determining its convergence. From the figures we also see the
structured variational EM is closer to MCEM than the fully factorial
variational EM. Therefore we use this method in the
remainder of experiments reported\footnote{
We also tested the full-factorial version, as well as 
the possibility of obtaining maximum a posteriori estimates for $u_n$
by conjugate gradients optimisation. Both have been found 
inferior to the structured variational EM approach, on the synthetic data
sets tested, both in terms of their accuracy of detecting genuine outliers, and 
their clustering accuracy rates evaluated against the true cluster labels.}.

\subsection{Assessing the accuracy of detecting genuine outliers}

To see how well can we detect genuine outliers,
we start by carrying out a set of controlled experiments,
varying the extent of measurement errors. We use our synthetic data
sets and define five different measurement error levels:
The diagonal elements of the error variance matrix $\mf{S}$ will
range between [0,0.01], [0,0.1], [0,1], [0,10] and [0,100]
respectively.

We perform receiver operating characteristics (ROC)
analysis~\cite{fawcett04roc} to measure the performance. The area
under the ROC curve (AUC) gives us the probability that a genuine
outlier is detected. The MoT is employed as a baseline in our
comparisons, in two instances: i) MoT applied to the clean data
(which in real applications is not available) provides an idealized
upper limit; ii) MoT applied to the data contaminated with
observation errors provides a baseline against of which we measure
our improvements. Fig. \ref{noiselevel} summarizes the results
obtained. For each of the 5 error conditions, the mean and standard
deviation of the AUC values over repeated runs on 30 independent
realizations of the data are shown: The upper plot shows the
in-sample performance whereas the lower plot shows the out-of-sample
performance, i.e. the ability to detect genuine outliers in
previously unseen data from the same density model.
\begin{figure}[htbp]
      \includegraphics[width = 1.6in, height=1.8in]{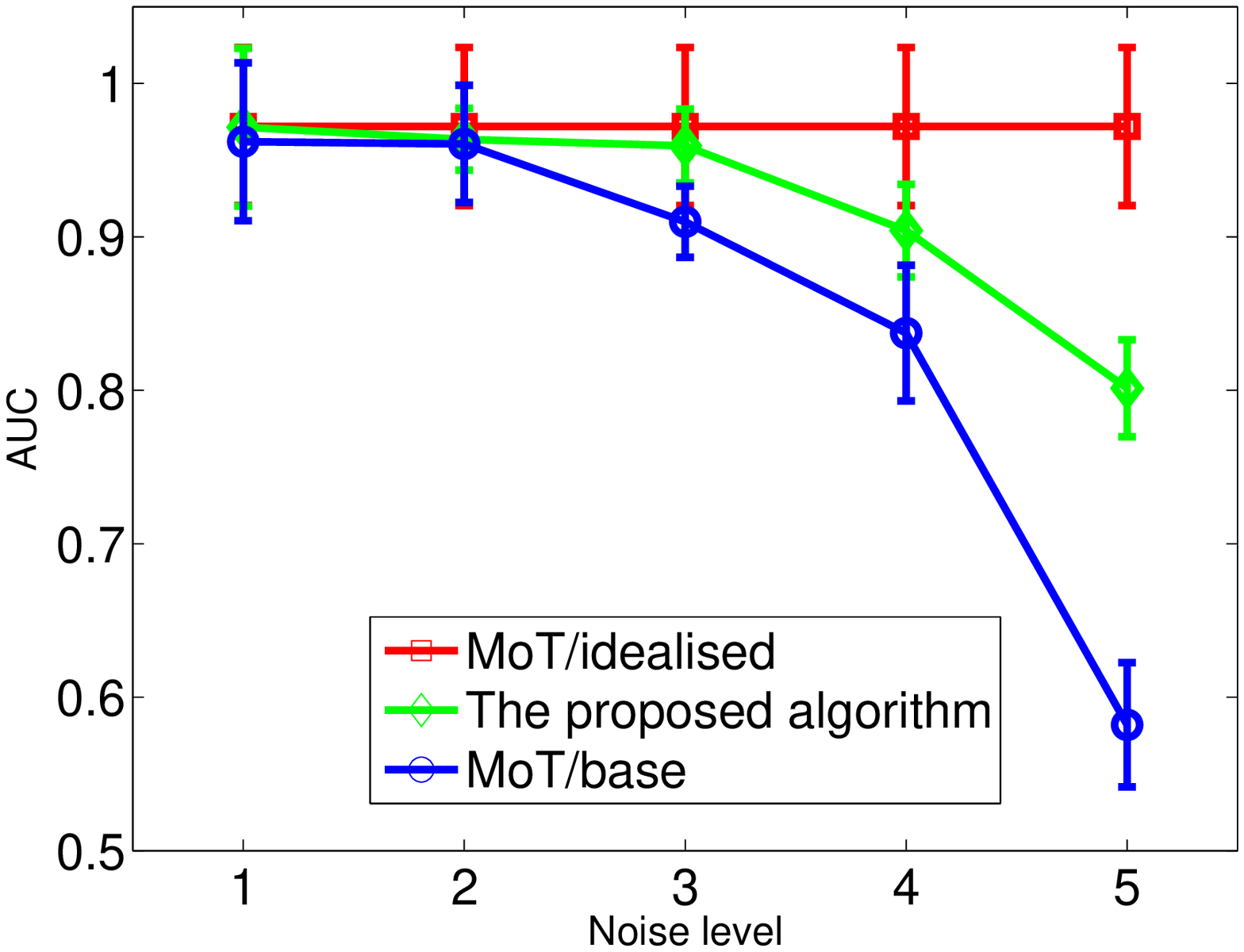}
      \includegraphics[width = 1.6in, height=1.8in]{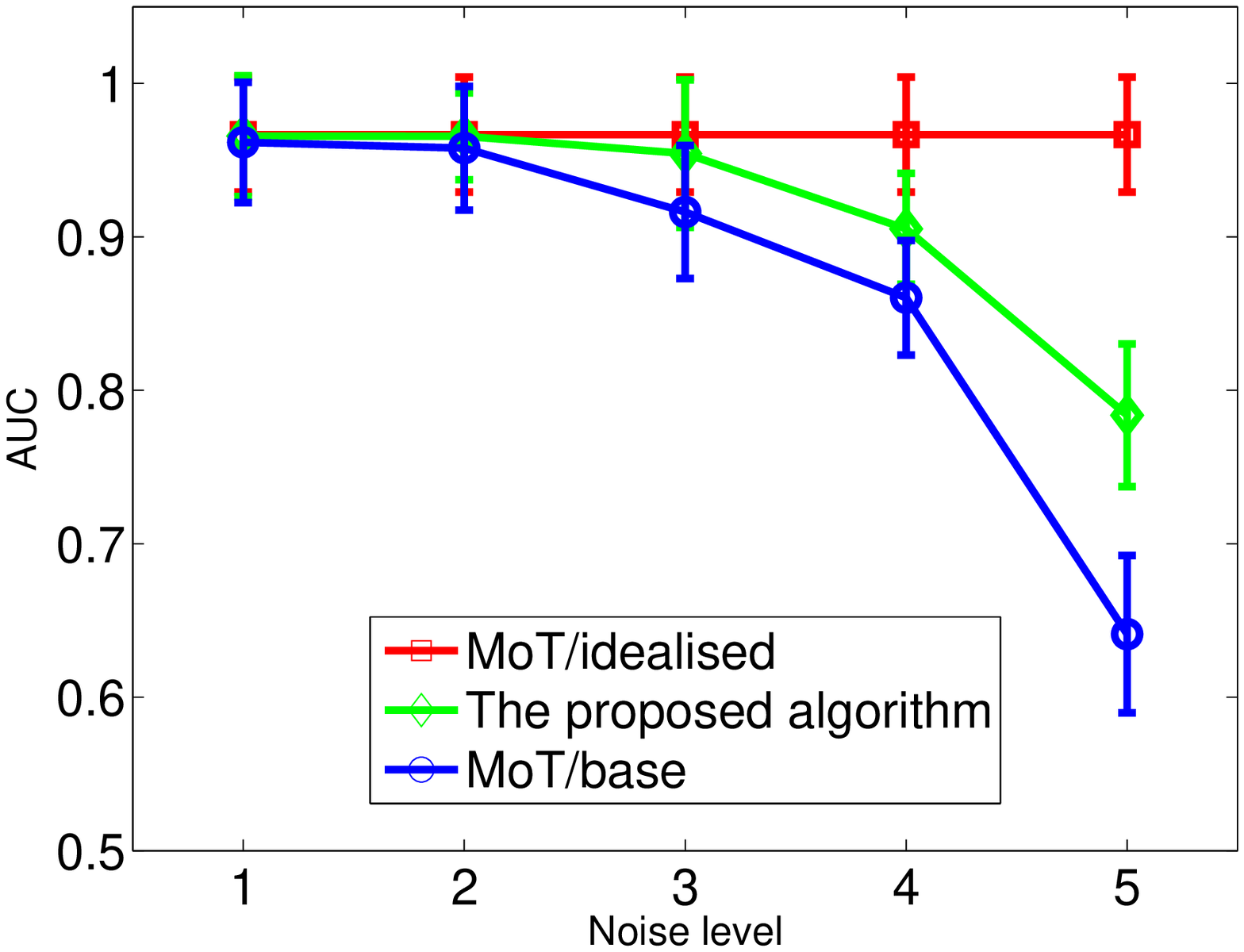}
\caption{Comparison of MoT/idealized, MoT/base and the proposed
algorithm on data sets with different levels of measurement error: in-sample (left)
and out-of-sample (right).}
    \label{noiselevel}
\end{figure}
The results are intuitive --- we see a systematic and increasingly
statistically significant improvement w.r.t. MoT/base, as the
measurement uncertainty increases, both on is-sample data and on
out-of-sample data.


In order to test our method further on data with a more realistic
underlying density, while still being able to evaluate the benefits
of using measurement error information in a controlled manner, 
we now apply our method on semi-synthetic data derived from real data,
the lymphography data set~\cite{lym}. Originally, the data
has four classes (148 data points in total and 18 dimensions), but
two of them are quite small (2 and 4 data records), so we consider
the two small classes as outliers. We added heteroscedastic Gaussian
noise with variances ranging between 0--0.1, to each observation, in
order to simulate errors.

The algorithms were run on 10 independent realizations of the
measurement errors and computed the average ROC curves \cite{fawcett04roc}
and associated average AUC. The in-sample
(93 data points) average AUC obtained by MoT/idealized is 0.9391, by
MoT/base it is 0.9005, whereas the proposed algorithm obtained
0.9555. The significance values of a t-test between MoT/idealized
and the proposed algorithm has been 0.39, while the value between
the proposed algorithm and MoT/base was $9.6\times 10^{-5}$. This
suggests that the proposed algorithm performs comparably to
MoT/idealized and significantly better than MoT/base in this
experiment. Moreover, the out of sample (55 data points) performance
has also been of the same quality and this is
shown in Fig. \ref{roc5}. We can conclude therefore, that knowledge of
measurement errors is useful and can be exploited with the use of
our approach to achieve a more accurate detection of genuine
outliers.
\begin{figure}[htbp]
  \begin{center}
     \includegraphics[width = 2.8in]{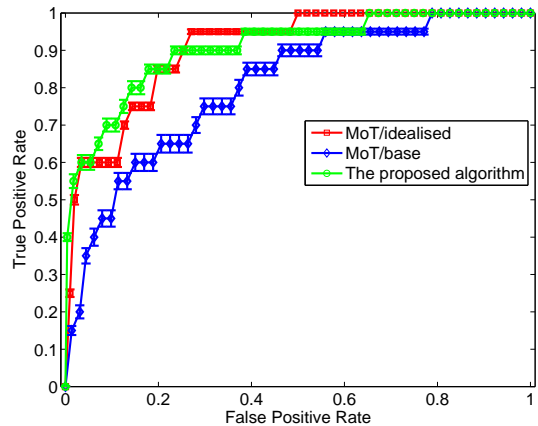}
\caption{Out of sample average ROC curves of MoT/idealized, MoT/base and the
proposed algorithm on the lymphography data set. The error bars represent
one standard deviation.}
    \label{roc5}
  \end{center}
\end{figure}

\subsection{Application to Detecting high-redshift quasars from the SDSS quasar catalogue}

In astrophysical measurements, there is no error-free situation \cite{Taylor}
since the objets observed are too far away. 
The measurement errors are known for each feature and each object, though
the error-free data is not accessible. Unlike in the previous sections, 
a validation against an absolute ground truth is therefore 
no longer possible. Nevertheless, the data set analyzed here is well-studied in
astrophysics, the SDSS quasar catalog \cite{vostat}, which provides
five magnitudes for a large number of quasars, representing their
brightness measured with five different filters
$u^\prime$ $g^\prime$ $r^\prime$ $i^\prime$ and $z^\prime$. From these,
to avoid bias with brightness, we construct four features, each related
to a color, by subtracting $r^\prime$ (which is the most reliably measured)
from each of the others.
In addition, spectroscopic redshift estimates are available
--- these are not used within the algorithm, but are useful to
derive a way of validating our results. The redshift is related to
the distance of the object from the Earth,
and very distant objects are rare. Given that with higher redshift,
the entire spectral pattern is systematically shifted towards
one end, there is physical reason for high redshift quasars to be
perceived as outliers in the overall density of quasars in the color
space. This observation has been exploited in a number of previous
studies for finding high redshift quasars in various
2D projections of the data~\cite{highestz}. However, a comprehensive
approach which both i) works in the multivariate feature
space and ii) takes principled account of the measurement errors has
not been available.

We apply our method to a sample of 10,000 quasars and compute the
AUC values against a varying redshift threshold. The resulting relationship
is shown in Fig. \ref{redshift}, for different choices of
$K$. The optimal order determined by MML was $K=2$, nevertheless,
from the figure we see a remarkable robustness w.r.t. this choice.
The y-coordinate of each point on these curves indicates the
probability of detecting quasars of redshift greater than its
x-coordinate. This plot shows clearly that our
principled method in four-color space, using errors, can identify as
outliers an overwhelming fraction of quasars already at a redshift of
2.5 (or higher), whereas the 2D projection methods, e.g. \singleemcite{highestz}, can
manage only those with $z\!>\!3.5$, which are extremely rare, and
obvious from naive projections. By being able to identify the latter
category, when the SDSS galaxy catalogue is complete with four-color
magnitudes, our method promises to retrieve an order of magnitude more
interesting high-redshift quasars than existing methods would.
\begin{figure}[htbp]
\begin{center}
\includegraphics[width = 2.8in]{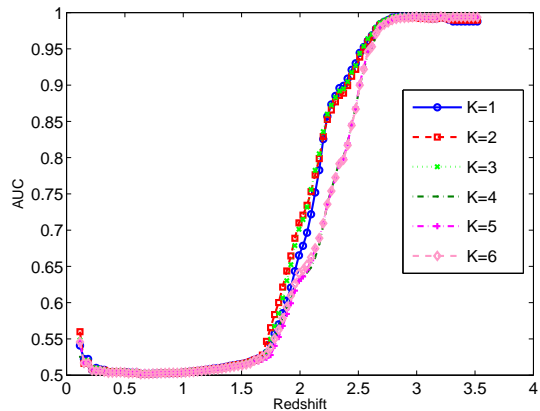}
\caption{AUC versus possible redshift thresholds. 
}\label{redshift}
\end{center}
\end{figure}

\section{Conclusions}\label{conclusion}

We proposed a robust mixture model for multivariate data that includes error
information. This was achieved by employing composite densities, designed to
infer peculiarity that is not due to errors.
We derived a structured variational EM algorithm for inference and parameter
estimation, which in the zero limit of the measurement errors
reduces to maximum likelihood estimation of $t$-mixtures. Assessment of the
variational scheme employed has shown it to be closer to the `ground truth' MCEM
than a fully factorial approximation scheme. Empirical results of a set of
controlled experiments have shown a systematic and statistically significant
improvement in terms of
correct outlier detection rates in high measurement uncertainty conditions, as a
result of appropriately incorporating knowledge about the
measurement errors. Finally, a real application of our method to
detecting peculiar, high-redshift quasars from the SDSS photometric
quasar catalogue was demonstrated.
Further work may concern extensions to robust projection models \cite{Outlier:arc&del}
and investigating ways of including an interactive visual
element into the analysis of outliers for data with error information.

{\small \bibliography{mybib_icml}}
\bibliographystyle{mlapa}

\end{document}